\documentclass[aps,prb,reprint,unsortedaddress,superscriptaddress,showpacs]{revtex4-1}
\usepackage{amsmath,amssymb,bm,graphicx}

\begin{document}

\title{Charge transport in pn and npn junctions of silicene}

\author{Ai Yamakage}
\affiliation{Department of Applied Physics, Nagoya University, Nagoya 464-8603, Japan}

\author{Motohiko Ezawa}
\affiliation{
Department of Applied Physics, University of Tokyo, Tokyo 113-8656, Japan
}
\author{Yukio Tanaka}
\affiliation{Department of Applied Physics, Nagoya University, Nagoya 464-8603, Japan}

\author{Naoto Nagaosa}
\affiliation{
Department of Applied Physics, University of Tokyo, Tokyo 113-8656, Japan
}
\affiliation{
RIKEN Center for Emergent Matter Science, ASI, RIKEN, Wako, Saitama 351-0198, Japan
}

\date{\today}

\begin{abstract}
We investigate charge transport of pn and npn junctions made from silicene, Si analogue of graphene. 
The conductance shows the distinct gate-voltage dependences peculiar to the topological and non-topological phases, where the topological phase transition is caused by external electric field.
Namely, the conductance is suppressed in the np regime when the both sides are topological, while in the nn regime when one side is topological and the other side is non-topological.
Furthermore, we find that the conductance is almost quantized to be 0, 1 and 2. 
Our findings will open a new way to nanoelectronics based on silicene.
\end{abstract}

%73.23.-b 	Electronic transport in mesoscopic systems
%72.80.Vp	Electronic transport in graphene
%73.40.-c	p-n junctions
\maketitle

\section{Introduction}

Technology fabricating one-atom thick systems has been rapidly developing since the appearance of monolayer honeycomb carbon (graphene).\cite{novoselov04}
Recently, silicon analog of graphene (silicene) \cite{GLay,Takamura,Kawai} has been synthesized and attracts much attention.
The low-lying excitations of the monolayer honeycomb systems are Dirac fermions.
Due to spin-orbit interaction (SOI), the Dirac fermions become massive, i.e., the energy band has a gap.
These massive Dirac fermion systems lead to a quantum spin Hall (QSH) insulator, which is originally proposed in graphene.\cite{kane05a,kane05b}
However, SOI of graphene is tiny so that the QSH effect in graphene has not been experimentally observed.
SOI of silicene is, in contrast, thousand times larger than that of graphene,\cite{Min, Yao}
which makes experimentally accessible QSH effects\cite{liu11prl}.

Transport properties of Dirac fermions show various anomalous behaviors.
A prominent feature is the Klein tunneling\cite{klein29}. Graphene heterojunctions exhibit perfect transmission through the barrier at normal incidence regardless of the barrier characteristics.
The origin of the Klein tunneling is the absence of the backscattering due to the pseudospin
conservation.\cite{katsnelson06, beenakker08, ando98}
The perfect transmission is protected by time--reversal symmetry.
Actually,  signature of Klein tunneling has been observed in graphene.\cite{huard07, shytov08, young09}
The systems supporting Dirac fermions can exhibit unique charge and spin transport.\cite{saito07, sonin09, yokoyama09, bercioux10, bai10, ingenhoven10, rataj11, bai11, bai11PhysicaE, niu11, guigou11, liu12, esmaeilzadeh12, tian12,tian12EPJB, rothe12, prada13} 
They have a potential to provide us with new electronics and spintronics devices.

Among such Dirac fermion systems,
 silicene has another advantage:
 the band gap is controllable by applying an external electric field\cite{ezawa12njp} owing to the buckling structure.\cite{takeda94}
 A bilayer graphene also has an electric-field-tunable band gap\cite{ohta06,mccann06,mccann06b,oostinga08,zhang09}.
 Silicene, differently from a bilayer graphene, shows a topological phase transition by tuning the band gap.
If one applies an electric field whose energy is stronger than SOI, the topological phase transition occurs from the QSH to non-topological insulators.
It is also intriguing that 
silicene realizes various topological insulators by exchange interaction,\cite{EzawaQAH} photo-irradiation\cite{EzawaPhoto} and anti-ferromagnetic order.\cite{EzawaExM}
% Rich properties of silicene will lead to future electronic devices.

% On the other hand,
% it is known that a Dirac fermion shows a
% Klein tunneling, especially, a perfect transmission in the massless case.\cite{Klein}
% It is difficult to observe such a tunneling for elementary particles, but it is possible in a graphene.
%
%
% There is a perfect tunneling in monolayer graphene, whereas there is a perfect reflection in bilayer graphene.
%
% An experimental verification of Klein tunneling in graphene opens a relativistic physics in condensed matter physics\cite{katsnelson06}.

These characteristics could be useful for device applications.
The most fundamental electronic device is a field-effect transistor (FET). 
FET made by silicene has an advantage that it has a large band gap due to SOI compared with graphene which is a zero gap semiconductor.
In addition, the tunable band gap and the topological phase transition of silicene by external electric field may lead to a new feature for FETs.
% Silicene is compatible to conventional silicon device technology since they are composed of the same atoms.

% However there has been no proposal on electronic devices based on silicene, which is crucial to future application of silicene. 
% The most fundamental electronic device is a field-effect transistor, where the conductance can be externally controlled by applying a gate voltage.

% Effects due to spin-orbit (SO) interaction have also been investigated by introducing it into graphene.
% However, the magnitude of the SO interaction is too small to detect its effect experimentally in graphene\cite{Min,Yao}.
% Recently, the honeycomb structure of silicon atoms named silicene has been experimentally synthesized\cite{GLay,Takamura,Kawai}.
% Silicene has a SO interaction thousand times larger than that of graphene, which naturally realizes quantum-spin Hall effects\cite{liu11prl}.

Charge transport properties of a silicene nanoribbon has been studied.\cite{ezawa13}
Spin transport has been also studied in a bulk silicene junction under Zeeman field.\cite{tsai13}
On the other hand, we focus on the charge transport in the bulk silicene. 
In this paper, we analyze the transport properties of pn and pnp junctions made of silicene. 
Controlling the conductance by tuning the gate voltage,
we find that i) the gate-voltage dependences of conductance in the topological and non-topological phases are quite distinct, and ii) the conductance is almost quantized to be 0, 1 and 2. 
The former is an evidence for the existence of the topological phase.
The latter enables us to use silicene as a FET with almost quantized three values of conductance.
Our results will open a new way to future nanoelectronics.

This paper is composed as follows. In Sec. \ref{bulk}, we review the bulk properties of silicene. In Sec. \ref{pnsec}, we investigate the pn junction of silicene. First we calculate the conductance for the normal incident case with and without the Rashba interaction. 
% We show the effects of the Rashba interaction is negligible. 
Next we calculate the obliquely incident case. 
The conductance is almost the same but smeared compared with that of the normal incident case.
In sec. \ref{pnpsec}, we investigate a pnp junction of silicene. 
Section \ref{summary} is devoted to discussions.

\section{Bulk  properties}
\label{bulk}

First, we show the bulk  properties of silicene.
The Hamiltonian of a silicene in the vicinity of the K and K$'$ points reads\cite{liu11prl,liu11,ezawa12njp}
\begin{align}
 H(\bm k)
 &= \hbar v_{\rm F} (k_x \tau_x - k_y \tau_y \eta_z)
 - \lambda_{\rm SO} \tau_z \sigma_z \eta_z
 \nonumber \\ & \quad
 + a \lambda_{\rm R} \tau_z (k_x \sigma_y - k_y \sigma_x) \eta_z
%  \nonumber \\ & \quad
 + \ell E_z  \tau_z,
 \label{hbulk}
\end{align}
where 
$\sigma_i$, $\tau_i$, and $\eta_i$ are the Pauli matrices for the spin ($\uparrow$ and $\downarrow$), sublattice (A and B sites) pseudospin, and valley (K and K$'$ points) spaces, respectively.
$a=3.86 \mathrm{\AA}$ and $2\ell = 0.46 \mathrm{\AA}$ denote the lattice constant and the perpendicular distance between A and B sites, respectively.

$\lambda_{\rm SO}$ is the intrinsic SOI coupling constant that triggers the topological phase transition from the non-topological to topological insulators (See below).
The sublattice-dependent Rashba SOI $\lambda_{\rm R}$ also appears due to the buckling structure of silicene.
And also, the mass term $\ell E_z \tau_z$ shows up because of the buckling structure with an external electric field $E_z$ along the $z$-axis.
Hereafter, we set $\hbar=1$.

\begin{figure}
\centering
\includegraphics{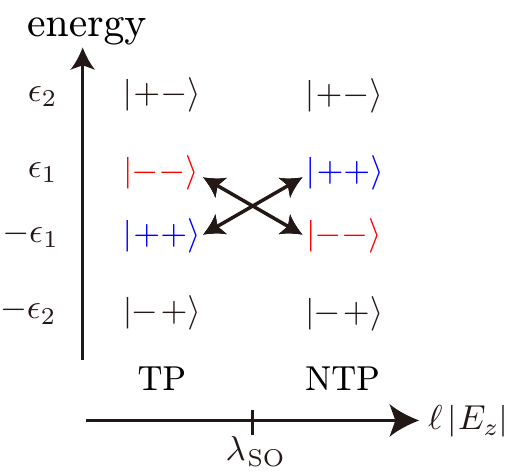}
\caption{Energy level scheme for the topological (TP) ($\ell \left| E_z \right| < \lambda_{\rm SO}$) and non-topological (NTP) ($\ell \left| E_z \right| > \lambda_{\rm SO}$) phases at the K point.
$\left| \alpha \beta \right \rangle$ denotes the eigenstate with $\tau_z = \alpha \, \mathrm{sgn} (E_z)$ and $\sigma_z = \beta \, \mathrm{sgn} (E_z)$.
$\epsilon_1 = \lambda_{\rm SO} + \ell \left| E_z \right|$, $\epsilon_2 = \left|\lambda_{\rm SO} - \ell \left| E_z \right| \right|$.
}
\label{energy0}
\end{figure}

We start with a review of the topological phase diagram of silicene.
The Dirac mass $m$ for the K point ($\eta_z=1$) in the bulk silicene is given by
%, which is defined by the Hamiltonian for the zero momentum, is given by
\begin{align}
 m = -\lambda_{\rm SO} \tau_z \sigma_z + \ell E_z \tau_z.
\end{align}
The system is a topological insulator for
$ \lambda_{\rm SO} > \ell |E_z|$,
while it is a non-topological insulator for $\lambda_{\rm SO} < \ell | E_z|$.
$ \lambda_{\rm SO} = \ell |E_z|$ is the critical point, where the energy gap closes. 
The sign-change of the mass term means the inversion between the conduction and valence bands.
Correspondingly, the directions of $\bm \tau$ and $\bm \sigma$ of the conduction and valence electrons change.
Namely, the sublattice and spin states are given by 
$\left| + - \right\rangle$, 
$\left| - - \right\rangle$,
$\left| + + \right\rangle$,
and 
$\left| - + \right\rangle$ in descending energy order for the topological phase.
On the other hand, the second and third states are interchanged for the non-topological phase, i.e., 
$\left| + - \right\rangle$, 
$\left| ++ \right\rangle$,
$\left| -- \right\rangle$,
and 
$\left| - + \right\rangle$.
Here $\left| \alpha \beta \right \rangle$ denotes the eigenstate with $\tau_z = \alpha \, \mathrm{sgn} (E_z)$ and $\sigma_z = \beta \, \mathrm{sgn} (E_z)$.
This energy level scheme is shown in Fig. \ref{energy0}.
As one switches off $E_z$, the two states with $\left| + - \right\rangle$ and $\left| - - \right\rangle$ (with $\left| + + \right \rangle$ and $\left| - + \right \rangle$) are  degenerated ($\epsilon_1=\epsilon_2$ in Fig. \ref{energy0}). 

\begin{figure}
\centering
\includegraphics{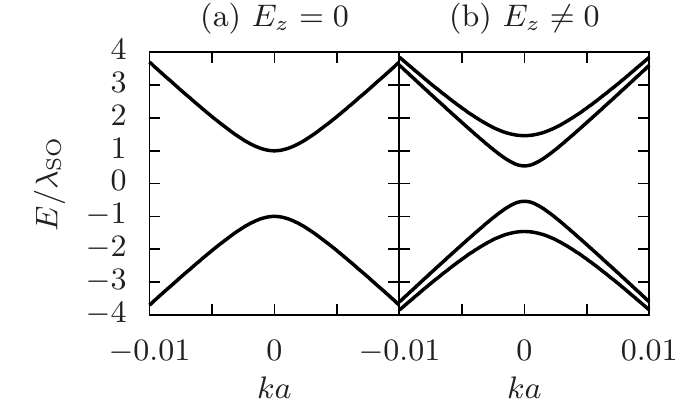}
\caption{Energy dispersions in the bulk silicene for $E_z=0$ (a) and $E_z = 0.5 \lambda_{\rm SO}$ (b).
The parameters of the system are taken as follows.
$v_{\rm F}/a = 1.4 \rm eV$, $\lambda_{\rm SO} = 3.9 \rm meV$, $\lambda_{\rm R} = 0.7 \rm meV$.
}
\label{bulkenergy}
\end{figure}
In addition, we show the energy dispersions for $E_z=0$ and $E_z \ne 0$ in Fig. \ref{bulkenergy}.  
The energy $E_\pm(\bm k)$ in the bulk is obtained to be
\begin{align}
 E_\pm^2(\bm k) = v_{\rm F}^2 k^2 + 
\left(\pm \sqrt{\lambda_{\rm SO}^2 + a^2 \lambda_{\rm R}^2 k^2} + \ell E_z
\right)^2,
\end{align}
with $k = (k_x^2+k_y^2)^{1/2}$.
The corresponding eigenvector $\bm u_{\pm}(\bm k; E_z)$ is also obtained analytically (Appendix \ref{wavefunction}).
The energy bands are doubly degenerated for $E_z=0$ due to the inversion
[$\tau_z \sigma_z H(\bm k) \sigma_z \tau_z = H(-\bm k)$]
 and  time-reversal 
[$\tau_y \sigma_x H^*(\bm k) \sigma_x \tau_y = H(-\bm k)$]
symmetries 
defined within each valley.
In contrast, there is no spin-degeneracy for $E_z \ne 0$ since $E_z$ breaks the inversion symmetry, which lifts the degeneracy at each k-point..

\section{Silicene pn junction}
\label{pnsec}

\begin{figure}
\centering
\includegraphics[scale=0.15]{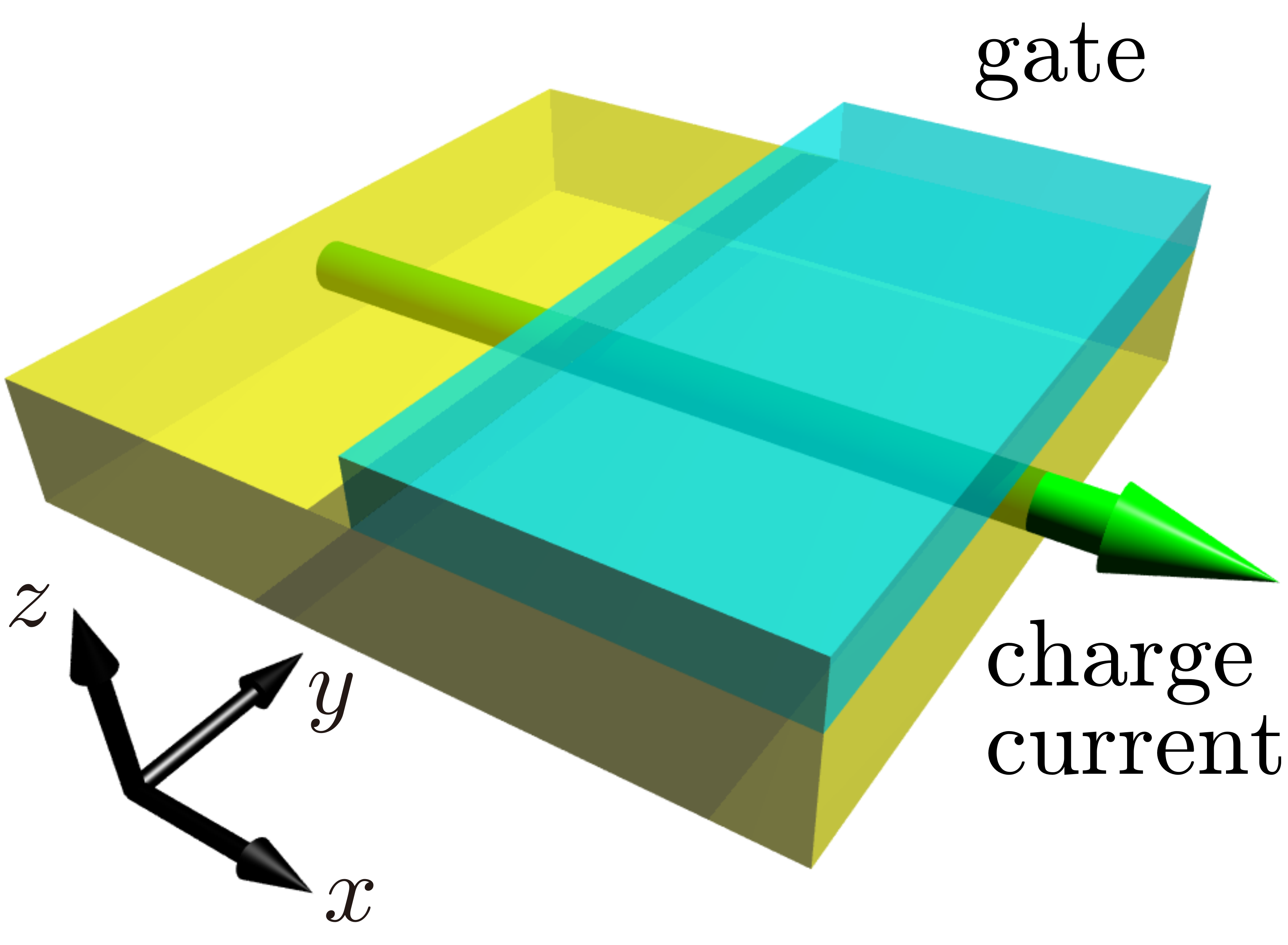}
\caption{Schematic of a pn junction.}
\label{pn}
\end{figure}

In this section, we investigate charge transport in a silicene pn junction, which is illustrated in Fig. \ref{pn}.

\subsection{Normal incident case}

\subsubsection{Formalism of the scattering problem}

Firstly, we investigate a normal incident of a pn junction of silicene for $k_y=0$. 
The Hamiltonian is given by
\begin{align}
 H(x) &= -i v_{\rm F} \partial_x \tau_x
 - \lambda_{\rm SO} \tau_z \sigma_z \eta_z
 + \ell E_z \theta(x) \tau_z
 \nonumber \\ & \quad
 + i a \lambda_{\rm R} \partial_x \sigma_y \tau_z \eta_z
  + V \theta(x).
\end{align}
Hereafter, we focus only on the K point ($\eta_z=1$). 
% The charge conductances from the K and K$'$ points are the same.
The same analysis is applicable to the K$'$ point.

We solve the scattering problem of the pn junction.
The calculation has been done by employing theories for graphene\cite{katsnelson06,beenakker08,cayssol09} and the Kane-Mele model.\cite{yamakage09,yamakage11}
In the incident side ($x<0$), an external electric field is not applied and hence the energy bands are doubly degenerated.
As a result,
there are two incident states for a fixed incident energy $E_{\rm F}$.
Wave function $\psi_{\pm}(x)$ of the scattering state 
with the incident energy being $E_{\rm F}$  has the form as
\begin{align}
\label{psil}
 \psi_{\pm}(-0) 
&= \bm u_{\pm}(k_{\rm I}; 0) + r_{\pm +} \bm u_+(-k_{\rm I}; 0) + r_{\pm -} \bm u_-(-k_{\rm I}; 0),
 \\
 \psi_\pm(+0) 
&= t_{\pm +} \bm u_+(q_+; E_z) + t_{\pm -} \bm u_-(q_-; E_z),
\end{align}
where the subscript of $\psi_\pm(x)$ denotes the spin state of the incident state $\bm u_\pm(k_{\rm I}; 0)$.
The first term $\bm u_\pm(k_{\rm I}; 0)$ of Eq. (\ref{psil})  denotes the incident state.
The other two terms correspond to the reflected states with the same ($\pm$) and different (spin-flip, $\mp$) spin states.
Momentum $k_{\rm I}$ of the incident electron is given by
\begin{align}
 k_{\rm I} = 
 \mathrm{sgn}
(
 E_{\rm F}
)
 \sqrt\frac
 {E_{\rm F}^2 - \lambda_{\rm SO}^2}
 {v_{\rm F}^2 + a^2 \lambda_{\rm R}^2}.
\end{align}
The sign of $k_{\rm I}$ is determined so that the group velocity of the incident state is positive.
Note that the incident state must be a propagating mode; $E_{\rm F}^2 > \lambda^2_{\rm SO}$.
Otherwise, the corresponding conductance is zero, by definition.
On the other hand,  momentum $q_\pm$ of the transmitted electron is obtained by solving the following equation;
\begin{align}
  \left( E_{\rm F} - V
\right)^2
 = 
v_{\rm F}^2 q_\pm^2 + 
\left(
 \pm \sqrt{\lambda_{\rm SO}^2 + a^2 \lambda_{\rm R}^2 q_\pm^2}
 + \ell E_z
\right)^2.
\end{align}
Since the group velocities of the transmitted electrons should be positive, 
the following relation is satisfied for the propagating mode;
\begin{align}
\mathrm{sgn} \left(q_\pm \right) = \mathrm{sgn} \left(E_{\rm F}-V \right).
\end{align}
For the evanescent mode, on the other hand, $\mathrm{Im} \, q_\pm > 0$ is satisfied.
And note that when 
$\left| E_{\rm F}-V \right| < \left| \lambda_{\rm SO}  - \ell \left| E_z \right|\right|$,
the system in $x>0$ becomes insulating, i.e., the resulting conductance vanishes.

The reflection and transmission coefficients $r_{\pm \pm}$ and $t_{\pm \pm}$ are obtained by solving the continuity condition at $x=0$.
Since the charge current is conserved, the following relation holds.
\begin{align}
 \left.
\frac{\partial H}{\partial (-i \partial_x)}
\right|_{x<0} \psi_\pm(-0)
 =
 \left.
 \frac{\partial H}{\partial (-i \partial_x)}
\right|_{x>0} \psi_\pm(+0),
\end{align}
where $\partial H/[\partial (-i \partial_x)]$ is the velocity operator.
The above relation is reduced to
\begin{align}
 \psi_{\pm}(-0) = \psi_{\pm}(+0).
\end{align}
Solving this, one obtains the reflection and transmission coefficients.
From the reflection coefficient $r_{\alpha\beta}$, the transmission probability $T_\pm$ is given by
\begin{align}
 T_\pm = 1 - \sum_{\beta = \pm} 
\left|r_{\pm \beta}
\right|^2.
\end{align}
% $T=2$ corresponds to a perfect transmission for both incident states.
Charge conductance $G$ in the normal incident case ($k_y=0$) is defined by
\begin{align}
 G = \frac{e^2}{h} (T_+ + T_-).
\end{align}

\begin{figure}
\centering
\includegraphics{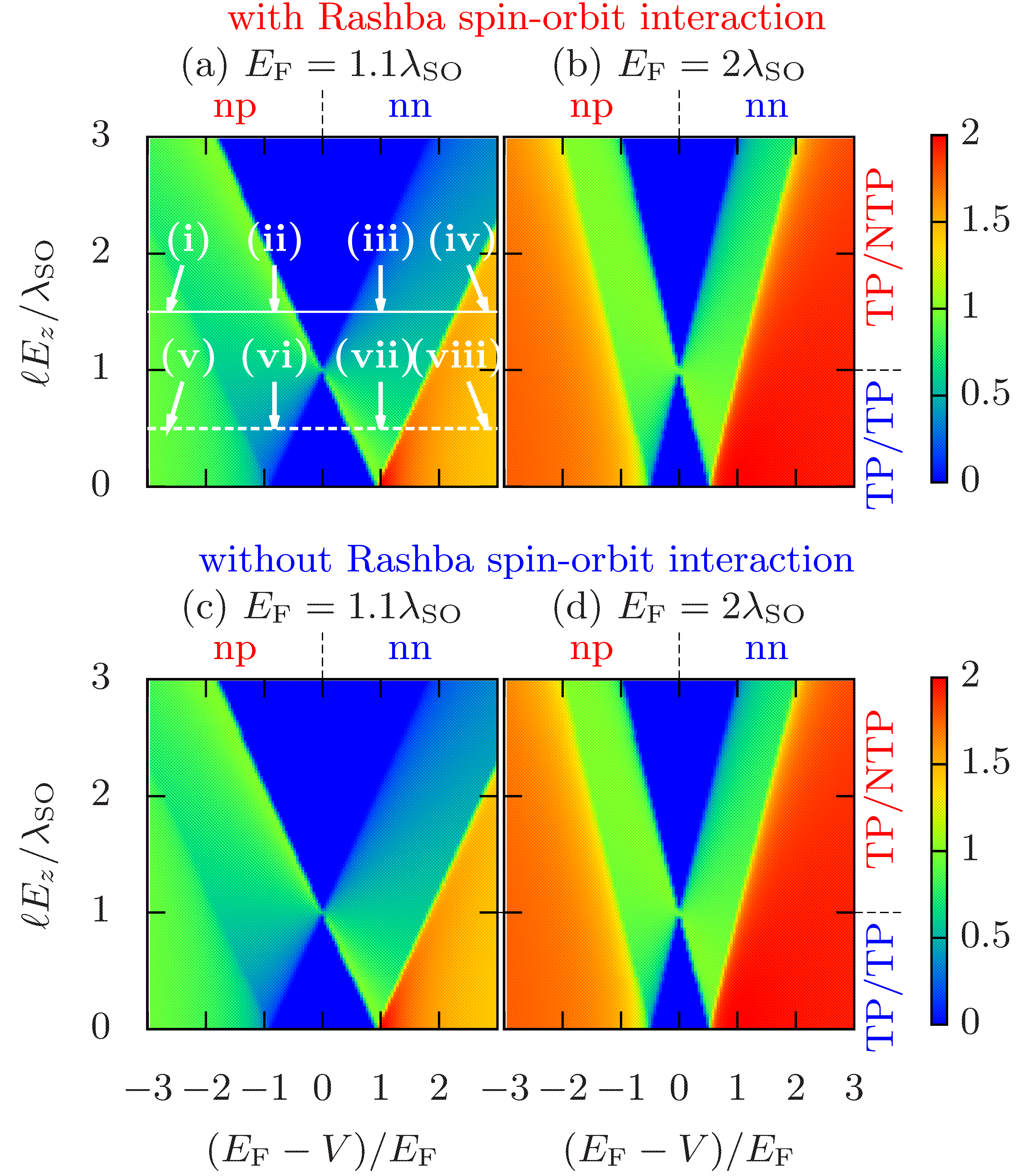}
\caption{Conductance in unit of $e^2/h$ for the normal incident case ($k_y=0$) in the presence [(a) and (b)] ($\lambda_{\rm R}/\lambda_{\rm SO} = 0.18$ corresponding to that for silicene) and absence [(c) and (d)] ($\lambda_{\rm R}=0$) of the Rashba SOI.
(a) and (c) [(b) and (d)] are the lightly (heavily) doped case as 
$E_{\rm F} = 1.1 \lambda_{\rm SO}$ 
($E_{\rm F} = 2 \lambda_{\rm SO}$).
The solid and dashed lines in (a) are located at $\ell E_z = 1.5 \lambda_{\rm SO}$ and $\ell E_z = 0.5 \lambda_{\rm SO}$, respectively.
(i)--(viii) are the representative points for which the conductance is shown in Fig. \ref{t1d}.
These are defined as follows.
(i) and (v): $V = 4\lambda_{\rm SO}$ (np and the double-channel regime).
(ii) and (vi): $V = 2\lambda_{\rm SO}$ (np and the single-channel regime).
(iii) and (vii): $V = 0$ (nn and the single-channel regime).
(iv) and (viii): $V = -2 \lambda_{\rm SO}$ (nn and the double-channel regime).
}
\label{g1D}
\end{figure}

% \subsubsection{results on transmission probability}
\subsubsection{Charge transport asymmetry in the nn and pn regimes}

We show results on the conductance of the normal incident case ($k_y=0$) in Fig. \ref{g1D}.
The horizontal axis is  $(E_{\rm F}-V)/E_{\rm F}$, where $E_{\rm F}-V$ corresponds to the Fermi energy in $x>0$ measured from the charge neutrality point.
The vertical axis is $\ell E_z/\lambda_{\rm SO}$.

Note that $E_z$ and $V$ are not actually independent of each other since both of them are induced by the gate electric field.
Therefore, the conductance along a curve in the $(E_{\rm F}-V, E_z)$ plane of Fig. \ref{g1D} is realized in the actual pn junction.
The relation between $E_z$ and $V$ depend on the substrate. It is worthwhile to investigate the general conductance formula depending on $E_z$ and $V$.

Only the region of $\ell E_z/\lambda_{\rm SO}>0$ is shown since the transmission probability is symmetric with respect to $E_z=0$ (See Appendix \ref{symmetry}).
A pn junction with two doped topological insulators (TP/TP) is realized for $\ell |E_z| < \lambda_{\rm SO}$. 
On the other hand, that with doped topological and non-topological (TP/NTP) insulators is realized for $\ell |E_z| > \lambda_{\rm SO}$.
Clearly seen from Fig. \ref{g1D}, there is no qualitative difference between the conductances with [Figs. \ref{g1D}(a) and (b)] and without [Figs. \ref{g1D}(c) and (d)] the sublattice-dependent Rashba SOI $\lambda_{\rm R}$.
This is because $\lambda_{\rm R}$ for silicene ($\lambda_{\rm R}/\lambda_{\rm SO}=0.18$) is weak and furthermore vanishes at the K and K$'$ points.

For $\lambda_{\rm R}=0$, one can obtain a simple formula for the reflection coefficient.
The reflection coefficient $r_{\sigma}$ with $\sigma = \pm$ being the $z$-component of spin of the incident electron is given by
\begin{align}
 r_\sigma = \frac{1-X_\sigma}{1+X_\sigma},
\end{align}
with
\begin{align}
 X_\sigma = 
\sqrt{
\frac
{E_{\rm F} + \sigma \lambda_{\rm SO}}
{E_{\rm F} - \sigma \lambda_{\rm SO}}
\frac
{E_{\rm F} - V - \sigma \lambda_{\rm SO} + \ell E_z}
{E_{\rm F} - V + \sigma \lambda_{\rm SO} - \ell E_z}
}.
\end{align}
The conductance is given by $G = (e^2/h) (2 - \sum_\sigma |  r_\sigma |^2)$.
If $\lambda_{\rm SO} \ll |E_{\rm F}|$ and $\ell |E_z| \ll |E_{\rm F}-V|$, the corresponding $r_\sigma$ tends to zero, i.e., a perfect transmission occurs, which is known as the Klein tunneling in monolayer graphene.\cite{katsnelson06,beenakker08}

Here we go back to Fig. \ref{g1D}.
In the inner region of 
$\left| E_{\rm F}-V \right| < \epsilon_1 \equiv \left|  \lambda_{\rm SO}  - \ell \left| E_z \right| \right|$, 
the conductance vanishes since the transmitted side ($x>0$) is insulating.
In the central region of $\epsilon_1 < 
\left| E_{\rm F} - V \right| < \epsilon_2 \equiv  \lambda_{\rm SO}  + \ell \left| E_z \right|$, there is a single energy band at the Fermi level, hence the maximum value of resultant conductance is $e^2/h$.
On the other hand, in the outer region of
$\left| E_{\rm F} - V \right| > \epsilon_2$,
two energy bands are located at the Fermi level.
Here the conductance becomes larger (almost double) than that in the central region.
We refer to these regions as insulating, single-channel, and double-channel regimes, respectively.
This behavior originates from a peculiarity of silicene, i.e., the band gap and spin-split energy bands owing to SOI and electric-field effect in the buckling structure.
Graphene, in contrast, does not have SOI nor the buckling structure. 
The resulting conductance is always $2 e^2/h$.

\begin{figure}
\centering
\includegraphics{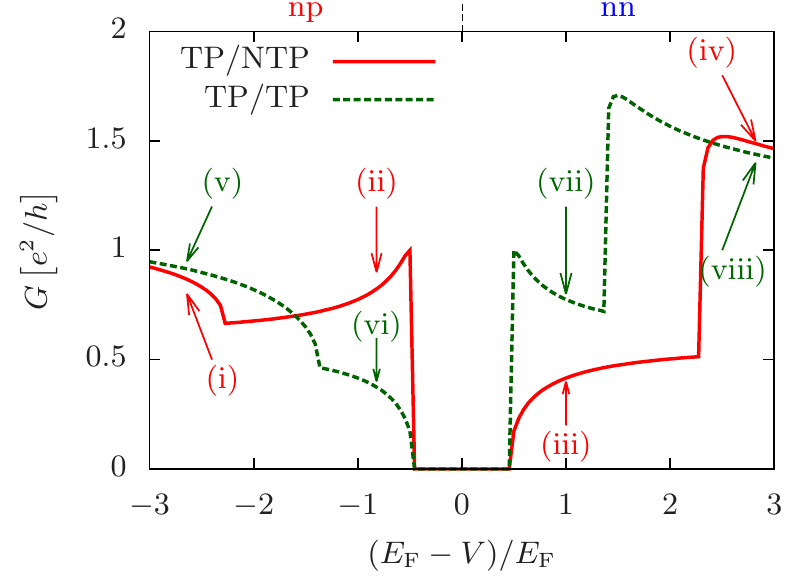}
\caption{Conductances for the TP/NTP ($\ell E_z = 1.5 \lambda_{\rm SO}$) and TP/TP ($\ell E_z = 0.5 \lambda_{\rm SO}$) junctions in the case of normal incidence. 
The parameters are the same as in Fig. \ref{bulkenergy}.
Cases (i)--(viii) correspond to those in Fig. \ref{g1D}(a).
}
\label{t1d}
\end{figure}

For the lightly doped case 
[$E_{\rm F}=1.1\lambda_{\rm SO}$, Figs. \ref{g1D}(a) and \ref{g1D}(c)],
one can see asymmetry of the conductance with respect to $\ell E_z = \lambda_{\rm SO}$.
To be more explicit we show the conductance as a function of $V$ for $E_z=1.5 \lambda_{\rm SO}$ (TP/NTP junction) and $E_z = 0.5 \lambda_{\rm SO}$ (TP/TP junction) in Fig. \ref{t1d}.
For 
the double-channel regime ($\left| E_{\rm F} - V \right| > \epsilon_2$),  
the transmission probabilities of the TP/NTP and TP/TP junctions are similar for both np [(i) and (v)] and nn [(iv) and (viii) case].
The transmission probability in the nn regime [(iv) and (viii)] is slightly larger than that in the np regime [(i) and (v)].
In contrast, for the single-channel regime 
($\epsilon_1 < \left| E_{\rm F} - V \right| < \epsilon_2$) 
[(ii), (iii), (vi), and (vii)], 
the conductances of the TP/NTP and TP/TP junctions are qualitatively different.
Namely, the conductance for the TP/NTP (TP/TP) junction in the np (nn) regime (ii) [(vii)]  takes a larger value than that in the nn  (np) regime (iii) [(vi)].
Note that the gate-voltage dependence of the conductance for the TP/NTP junction is distinct to that for the TP/TP junction.

This asymmetric behavior of conductance stems from the sublattice and spin states of the incident and transmitted electrons.
%%%%%%%%%%%%%%%%%
\begin{figure*}
\centering
\includegraphics{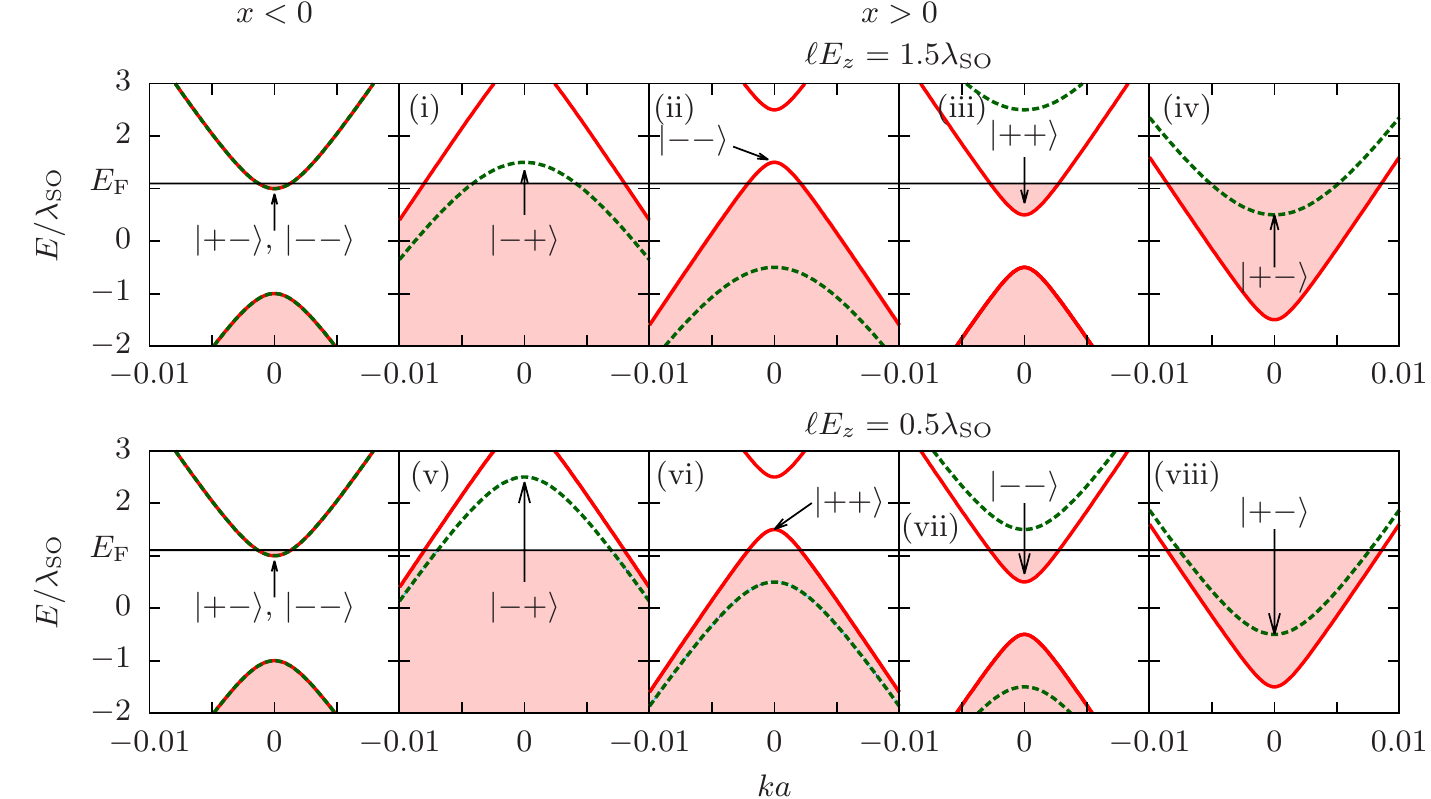}
\caption{Energy spectra for the incident ($x<0$) state with $\ell E_z = 0$ (the left-upper and left-lower panels), and for transmitted ($x>0$) states with $\ell E_z = 1.5 \lambda_{\rm SO}$ [(i)--(iv)] and with $\ell E_z =0.5$ [(v)--(viii)].
Cases (i)--(viii) corresponds to those in Fig. \ref{g1D}.
$|\alpha\beta \rangle$ with $\tau_z = \alpha \, \mathrm{sgn} \left( E_z \right)$ and $\sigma_z = \beta \, \mathrm{sgn} \left( E_z \right)$ denotes the sublattice and spin state for $k=0$.
}
\label{energy}
\end{figure*}
%
%
%
% \begin{figure}
% \centering
% \includegraphics{cnd_V.eps}
% \caption{Normalized conductance $G/G_0$ as a function of $E_{\rm F}-V$ for $\ell E_z = 0.5 \lambda_{\rm SO}$ (a) and $\ell E_z = 1.5 \lambda_{\rm SO}$ (b) with $E_{\rm F} = 1.1 \lambda_{\rm SO}$.
% These parameters are indicated by solid and dashed lines in Fig. \ref{g2d}(a).
% }
% \label{cnd_V}
% \end{figure}
Figure \ref{energy} shows the energy bands for $x<0$ and $x>0$.
The sublattice and spin states $\left | \alpha \beta \right\rangle$ for each energy band at $k=0$ are also denoted in Fig. \ref{energy}.
The incident states ($x<0$) with a positive energy
are approximately given by $\left| + - \right \rangle$ and $\left| -- \right\rangle$.
When the transmitted states is given by $\left| -- \right\rangle$ [(ii) and (vii)] or $\left| +- \right\rangle$ [(iv) and (viii)], the conductance is large, due to matching of the sublattice and spin states.
Especially, for $(E_{\rm F}-V)/E_{\rm F} \sim -0.5$ and $(E_{\rm F}-V)/E_{\rm F} \sim 0.5$, the transmission probabilities are unity since the sublattice and spin states of the incident and transmitted electrons coincide with each other. 
In contrast, when the transmitted state is given by the mismatched state $\left| ++ \right \rangle$ [(iii) and (vi)] or $\left| -+ \right\rangle$ [(i) and (v)], the corresponding conductance is suppressed.
Thus the matching/mismatching of the sublattice and spin states gives a larger/smaller conductance.

It is emphasized that the transmitted states $\left| ++ \right \rangle$ and $\left| -- \right \rangle$ are controlled by $E_z$.
As shown in Fig. \ref{energy0}, the two states $\left| ++ \right \rangle$ and $\left| -- \right \rangle$ are interchanged in the different topological phases, which is determined by $E_z$.
In other words, the conductance is well tuned by $E_z$ through changing the symmetry of the wave function.
The obtained results are summarized in Table. \ref{tab1}.
The same behavior has been observed in Ref. \onlinecite{yokoyama10} on the surface of a topological insulator with ferromagnets.
\begin{table}
\centering
\begin{tabular}{l|cc}
\hline
\hline
 & np & nn
\\
\hline
TP/NTP & large & small
\\
TP/TP & small & large
\\
\hline\hline
\end{tabular}
\caption{
Magnitudes of conductances in the silicene junctions for the single-channel regime 
($\epsilon_1 < |E_{\rm F}-V| < \epsilon_2$).
In the junction, the two cases are realized:
the gated region is nontopological (TP/NTP) and
topological (TP/TP).
}
\label{tab1}
\end{table}

As explained above, the conductance controlled by $E_z$ is determined by the matching of the sublattice and spin states between the both sides of the junction.
Therefore, this behavior does not appear in the heavily doped case ($\left|E_{\rm F}\right| \gg \lambda_{\rm SO}, \ell \left| E_z \right|$) [Figs. \ref{g1D}(b) and \ref{g1D}(d)], where the mass gap $\sim \lambda_{\rm SO}$ is negligible as compared to $E_{\rm F}$.
The conductance asymmetry peculiar to the topological phase can be expected for other topological insulators, provided that the system has a single Fermi surface in the pn junction.

\subsubsection{Heavily doped case}

From Figs. \ref{g1D} (b) and (d), the transmission probability is almost quantized to be 0 for the insulating case 
($|E_{\rm F}-V| < \epsilon_1$), to 1 for the single-channel regime 
($\epsilon_1 < |E_{\rm F}-V| < \epsilon_2$), and to 2 for the double-channel regime 
($ | E_{\rm F} -V | > \epsilon_2$).
This is a consequence of the Klein tunneling of Dirac fermions:
A massless Dirac fermion can tunnel through any barriers. 
Hence the normal incident transmission probability is unity.\cite{katsnelson06, beenakker08}
Although silicene has a finite energy gap, a perfect transmission approximately occurs for a heavily doped case, since the energy gap is effectively ignored as compared to the incident energy.
In contrast, a graphene pn junction always shows a perfect transmission, i.e., the value of conductance is always $2 e^2/h$.
Thus graphene cannot be used as a FET.

\subsection{Obliquely incident case}
\label{sec2dpn}

Next we turn to the case of finite $k_y$, which corresponds to an actual silicene pn junction.
The Hamiltonian of the two-dimensional system in the vicinity of the K point reads
\begin{align}
 H(x) &= -i v_{\rm F} \partial_x \tau_x 
 - k_y \tau_y
 - \lambda_{\rm SO} \tau_z \sigma_z 
 + \ell E_z \theta(x) \tau_z
 \nonumber \\ & \quad
 -  a \lambda_{\rm R} (-i\partial_x \sigma_y - k_y \sigma_x) \tau_z 
  + V \theta(x).
\end{align}
Here we assume translational invariance along the $y$-axis, i.e., the $y$-component of momentum $k_y$ is regarded as a parameter.
We  solve the scattering problem in the same way as in the previous section.

The normalized conductance $G/G_0$  is given by
\begin{align}
 \frac{G}{G_0} = \int_{-\pi/2}^{\pi/2} \frac{d\theta}{2} \cos \theta \, [T_+(\theta) + T_-(\theta)],
 \label{gg0}
\end{align}
with $T_\pm(\theta)$ being the transmission probability for an incident angle $\theta$ defined by $k_y = k_{\rm F} \sin \theta$ and for incident spin $\pm$.
In the case of perfect transmission ($T_\pm(\theta)=1$), the resulting conductance takes the value of $G = 2G_0$, where factor 2 means that the system has two incident states $\bm u_+$ and $\bm u_-$ with different spin states.
Here,
$G_0 =  (k_{\rm F} W / \pi) e^2/h$,
$k_{\rm F} = [(E_{\rm F}^2-\lambda_{\rm SO}^2)/(v_{\rm F}^2 + a^2 \lambda_{\rm R}^2)]^{1/2}$,
 and $W$ being the width of the system.
Also, Fano factor $F$, which corresponds to the shot noise-to-signal ratio, is given  by
\begin{align}
 F = \frac{G_0}{G} \sum_{\alpha = \pm} \int_{-\pi/2}^{\pi/2} \frac{d\theta}{2} \cos \theta \, T_\alpha(\theta) \left[ 1-T_\alpha(\theta) \right].
 \label{deffano}
\end{align}

\begin{figure}
\centering
\includegraphics{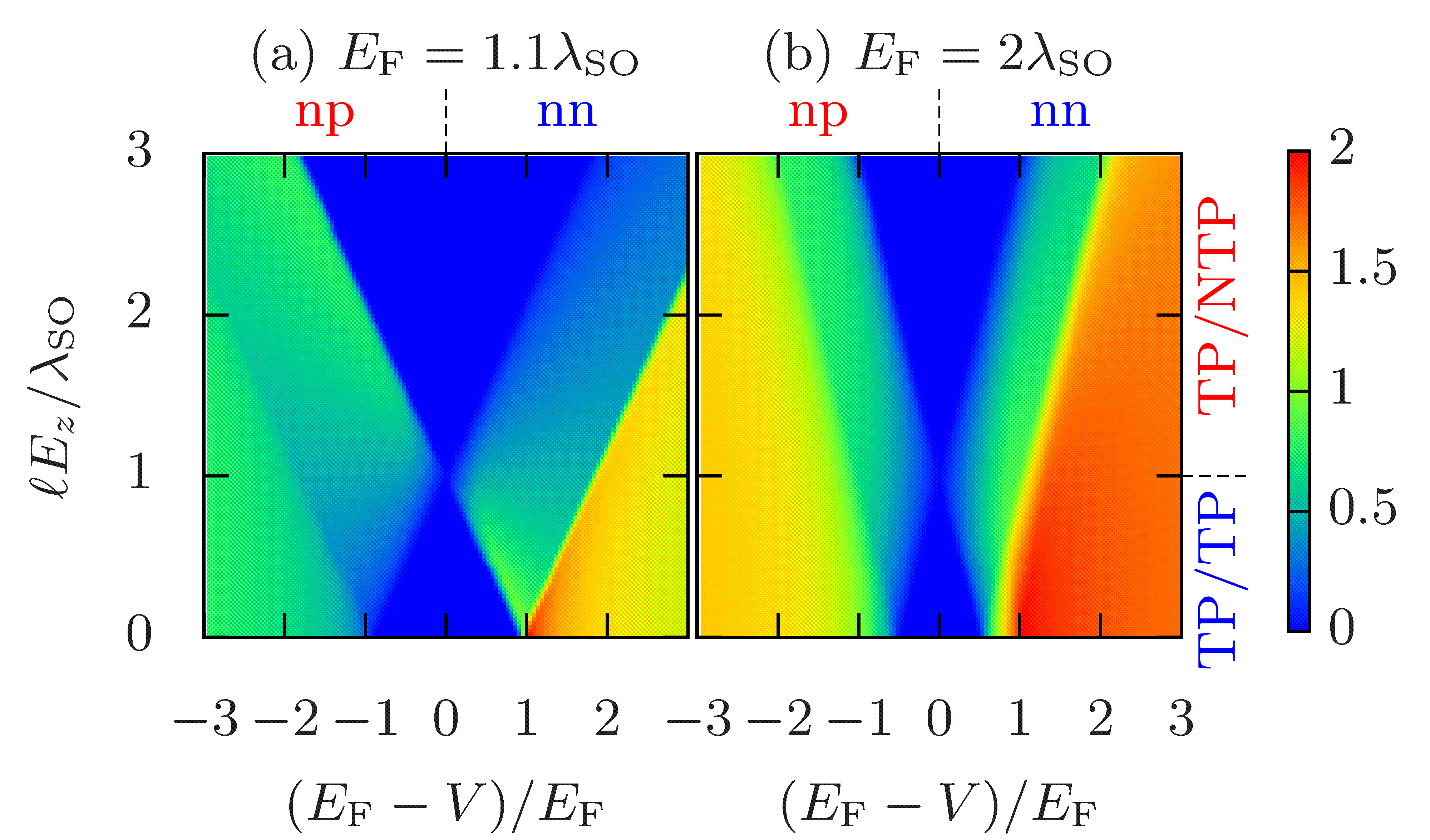}
\caption{Normalized conductance $G/G_0$ averaged over the incident angles for the lightly (a) ($E_{\rm F}=1.1 \lambda_{\rm SO}$)  and  the heavily (b) ($E_{\rm F}=2\lambda_{\rm SO}$)  doped cases. 
}
\label{g2d}
\end{figure}

Figure \ref{g2d} shows the normalized charge conductance $G/G_0$ as a function of  gate voltage [$(E_{\rm F}-V)/E_{\rm F}$] and electric field ($\ell E_z/\lambda_{\rm SO}$).
Obviously, the charge conductance  (Fig. \ref{g2d}) and the transmission probability of the normal incident case [Figs. \ref{g1D}(a) and \ref{g1D}(b)] are almost the same, except for the broadening of line shape.
This is because transport is determined basically by the normal incidence. 
An integral over the incident angle $\theta$ solely gives line broadening of the charge conductance from that for the normal incidence $T(\theta=0)$.

\begin{figure}
\centering
\includegraphics{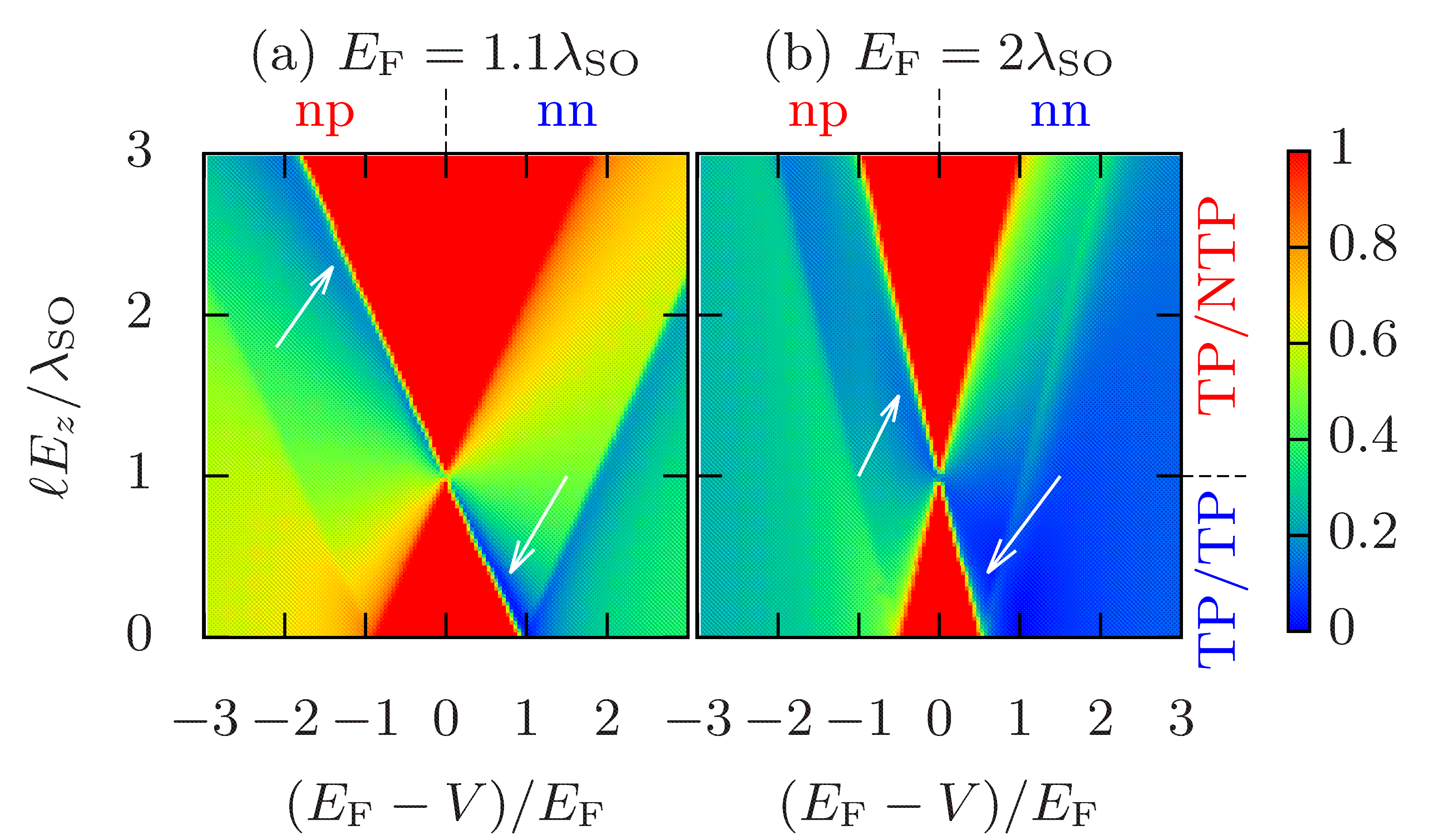}
\caption{Fano factor
for the lightly (a) ($E_{\rm F}=1.1 \lambda_{\rm SO}$)  and  the heavily (b) ($E_{\rm F}=2\lambda_{\rm SO}$)  doped cases. 
}
\label{fano}
\end{figure}

The Fano factor of the junction is shown in Fig. \ref{fano}.
Overall, the resulting Fano factor for the lightly doped case (a) is smaller than that for the heavily doped case (b).
Also,
the Fano factor is roughly given by the inverse of conductance, i.e.,
it takes a small (large) value when the corresponding conductance is large (small).
This behavior is realized if the shot noise power is almost independent of the parameters ($V$ and $E_z$).
On the other hand,
in the single-channel regime,
the Fano factor is strongly suppressed when the sublattice and spin states of the incident and transmitted electrons coincide with each other (denoted by the arrows in Fig. \ref{fano}).
Note that the Fano factor in the insulating region ($|E_{\rm F} - V| < \epsilon_1$) is obtained to be unity, although it is not well-defined for metal-insulator junctions because the corresponding conductance vanishes ($F \to 0/0$).
We have concluded $F=1$ in the insulating region since $F = 1$ has been obtained for the long junction limit of the npn junction, as discussed in the next section.

\section{silicene npn junction}
\label{pnpsec}

\begin{figure}
\centering
\includegraphics[scale=0.15]{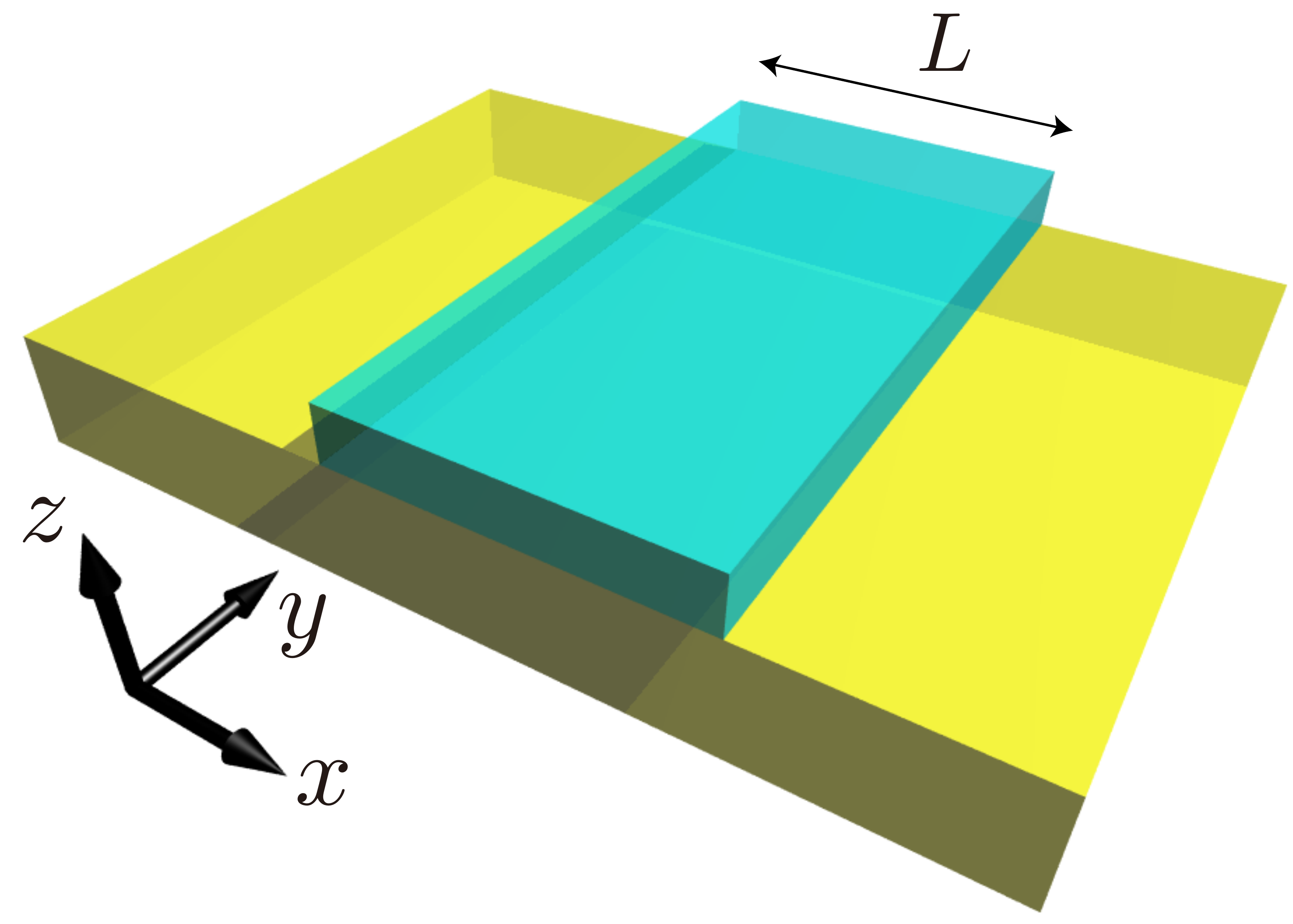}
\caption{Silicene npn junction. 
The center region (the length  $L$) is gated.
Charge current flows along the $x$-axis.}
\label{pnp}
\end{figure}

Next we investigate charge transport in a silicene npn junction, 
where electrostatic field is applied in $0<x<L$, which is illustrated in Fig. \ref{pnp}.
The scattering problem of the npn junction is solved in a manner similar to that of the pn junction.
The wave function $\psi_\pm(x)$ is given by
\begin{align}
 \psi_\pm(x=-0) 
&= \bm u_\pm(k_{\rm I}; 0) 
+ r_{\pm +} \bm u_+(-k_{\rm I}; 0) 
\nonumber\\ & \qquad
+ r_{\pm -} \bm u_-(-k_{\rm I}; 0),
 \\ 
 \psi_\pm(0<x<L) 
&= \sum_{i=1}^4 w_{\pm i} 
\bm u_{\alpha_i} (q_i; E_z) e^{i q_i x},
\\
\psi_\pm(x=L+0) 
&= 
t_{\pm +} \bm u_+(k_{\rm I}; 0) e^{i k_{\rm I} L}
\nonumber \\ & \qquad
+ 
t_{\pm -} \bm u_-(k_{\rm I}; 0) e^{i k_{\rm I} L},
\end{align}
where $q_i$ and $\alpha_i$ are solutions of $E_{\rm F}-V = E_{\alpha_i}(q_i,k_y)$.
Coefficients $r_{\pm \pm}$, $w_{\pm i}$, and $t_{\pm \pm}$ are obtained by solving the following boundary condition:
\begin{align}
 \psi_\pm(-0) &= \psi_\pm(+0),
 \\
 \psi_\pm(L-0) &= \psi_\pm(L+0).
\end{align}
The normalized conductance and the Fano factor are obtained by Eqs. (\ref{gg0}) and (\ref{deffano}), respectively.

\subsection{Normal incident case}

\begin{figure}
\centering
\includegraphics{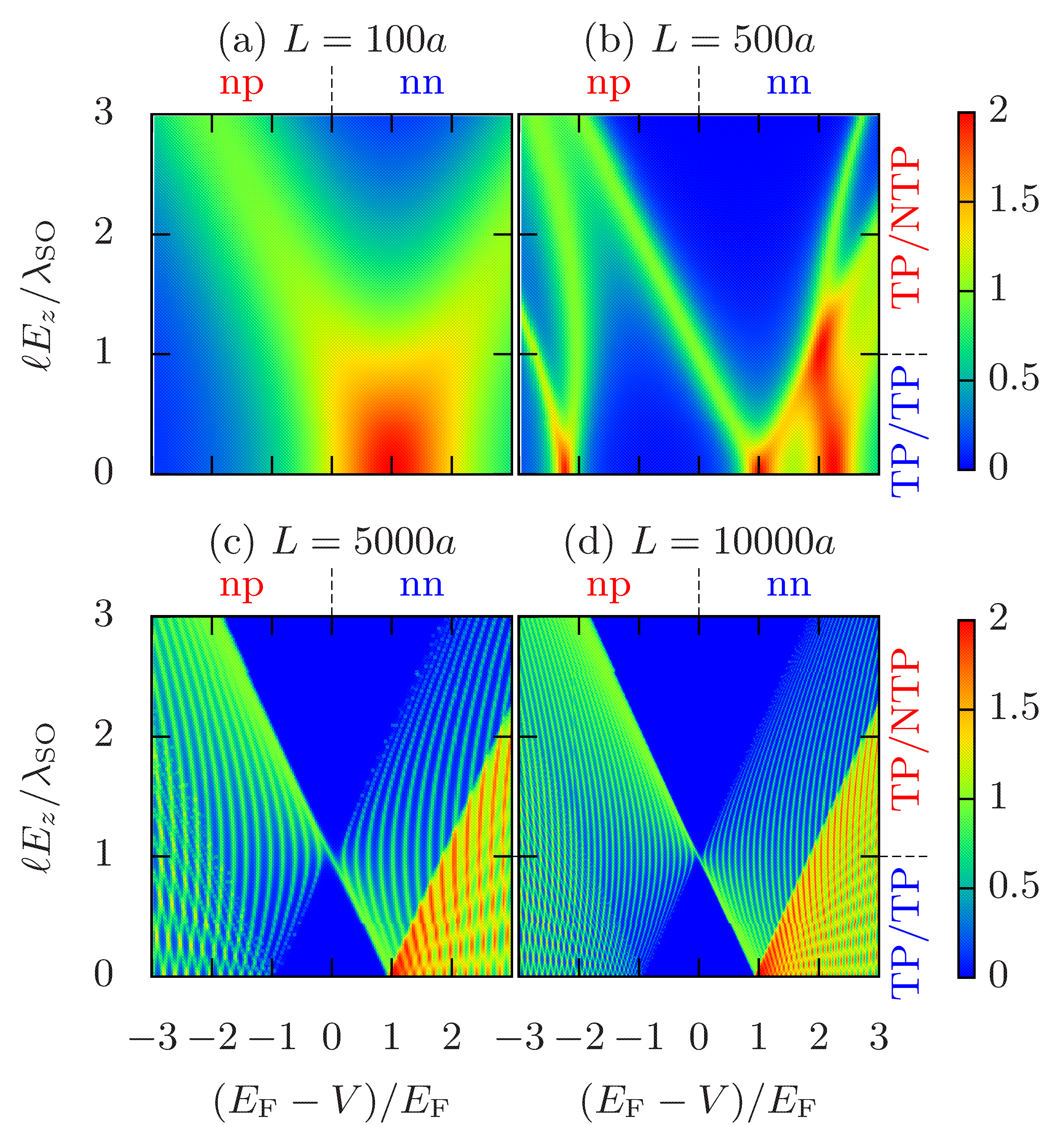}
\caption{Conductance in unit of $e^2/h$  for npn junction ($\theta=0$) in the normal incident case ($k_y=0$).
$L$ and $a$ denote the length of the gate in the junction and the lattice constant, respectively.
The incident energy is taken to be $E_{\rm F}=1.1 \lambda_{\rm SO}$.
}
\label{pnp1d}
\end{figure}

First we show the conductance for the normal incident case in Fig. \ref{pnp1d}, 
i.e., $T_+(0) + T_-(0)$.
A resonant tunneling occurs for $q_iL = 2n\pi, n \in \mathbb Z$ in a npn junction.
In the short junction limit ($L \to 0$), a perfect transmission always occurs even when the central region is insulating.
In a short but finite-length junction [Fig. \ref{pnp1d}(a)], the number of resonant peaks ($q_i = 2n\pi/L$) is still small (two peaks). On the other hand, the peak width is broad ($\sim 2 \lambda_{\rm SO}$) since the length of the junction is short so that the transmission probability is large. 
Thus, two broad resonant peaks appear in Fig. \ref{pnp1d}(a).
As one increases $L$, the number of resonant peaks increases and the peak width becomes narrower, as shown in Figs. \ref{pnp1d}(b) and \ref{pnp1d}(c). 
Finally, the transmission probability of the long junction ($L > 10000a$) [Fig. \ref{pnp1d}(d)] asymptotically converges to that of the pn junction [Fig. \ref{g1D}(a)].

\subsection{Obliquely incident case}

Next we show results on the obliquely incident case.
%%%%%%%%%%
\begin{figure}
\centering
\includegraphics{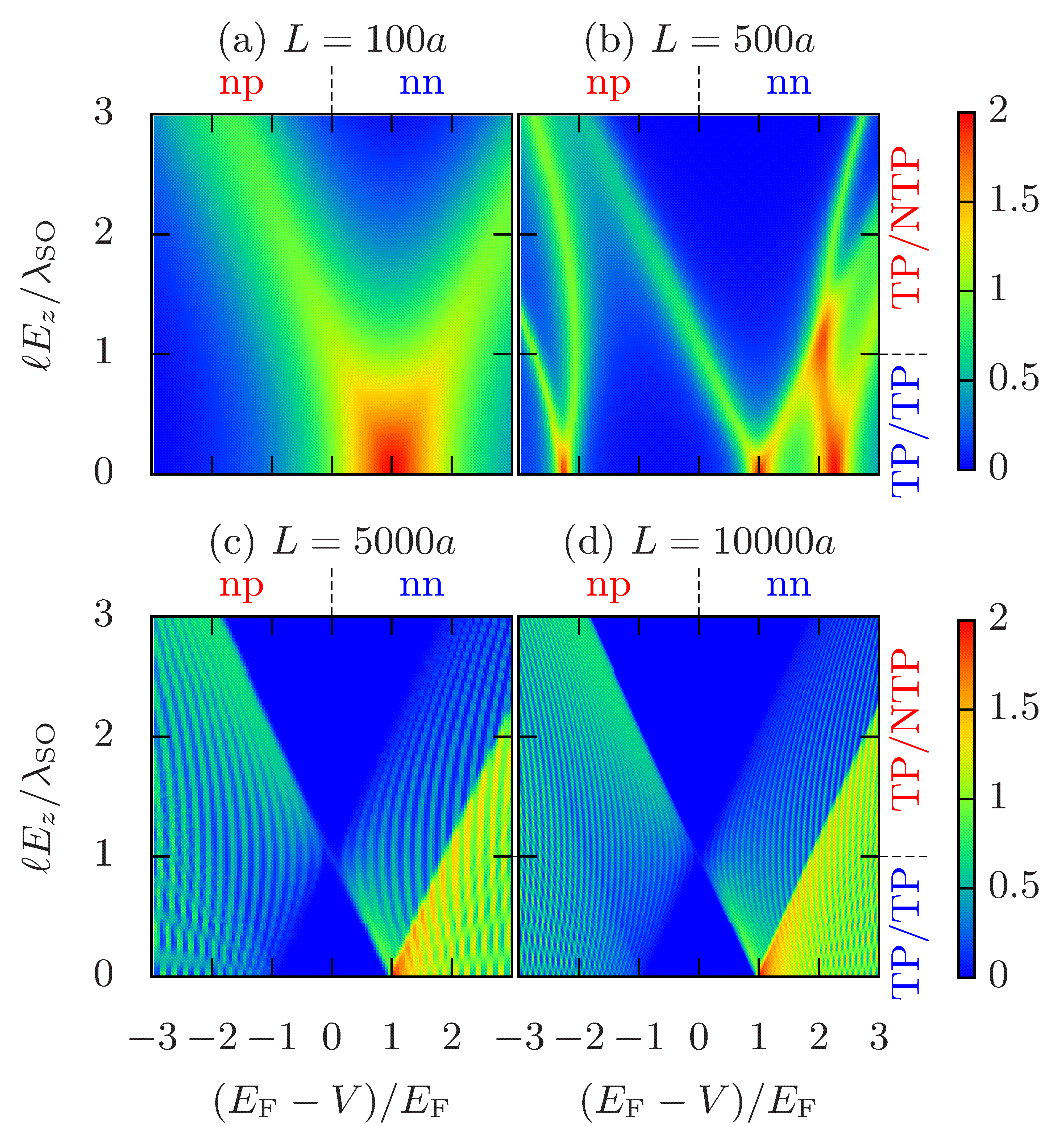}
\caption{Normalized Conductance $G/G_0$ in the npn junction for $E_{\rm F}=1.1 \lambda_{\rm SO}$.}
\label{pnp2d}
\end{figure}
%%%%%%%%%%%
The charge conductance is shown in Fig. \ref{pnp2d}.
As in the case of the pn junction discussed in Sec. \ref{sec2dpn}, integral over the incident angle entirely causes broadening of detail structures in the conductance. 
Namely, the conductance [Figs. \ref{pnp2d}(a)-(d)] is almost the same as that of the normal incident case [Figs. \ref{pnp1d}(a)-(d)].

\begin{figure}
\centering
\includegraphics{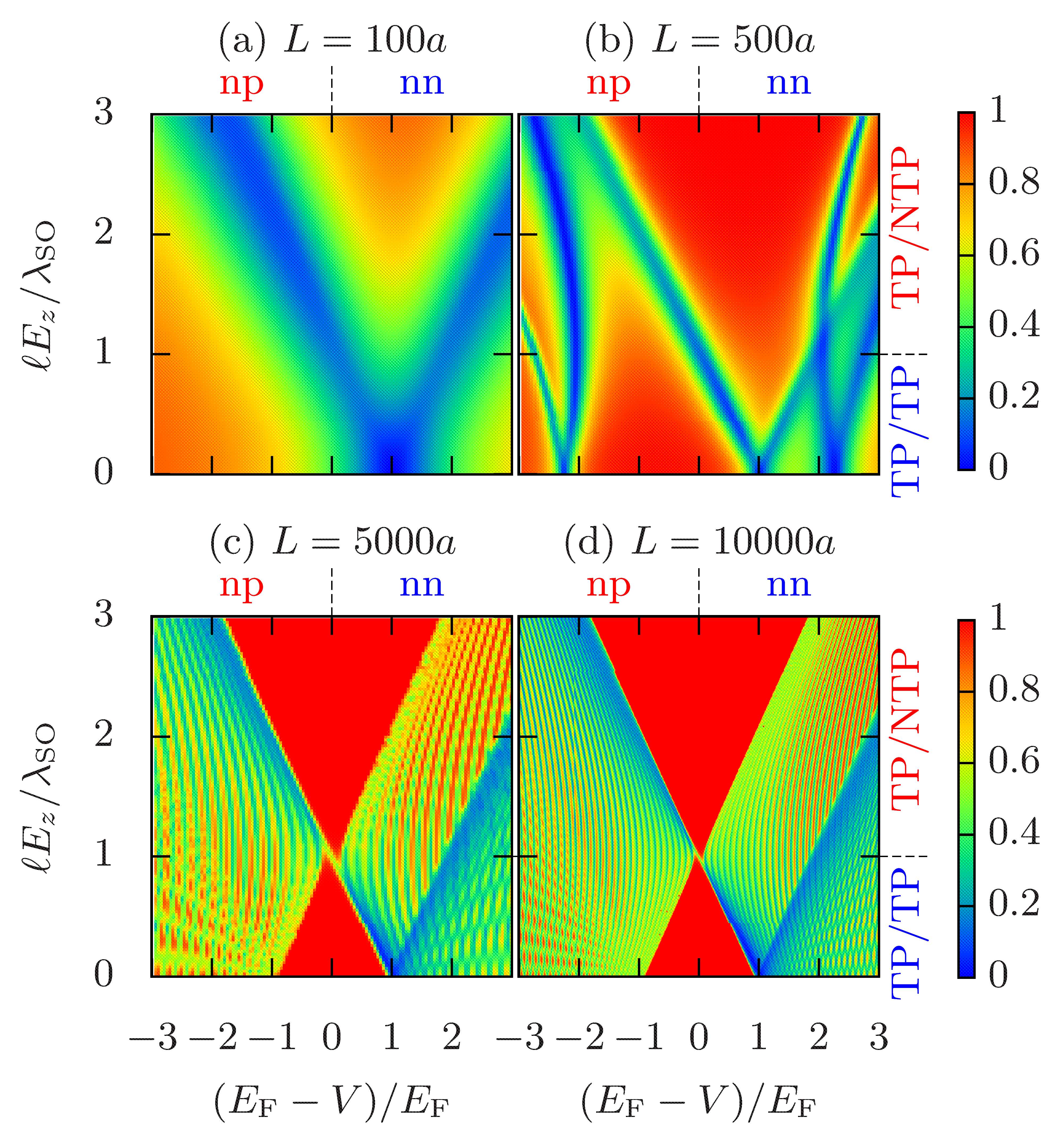}
\caption{Fano factor in the  npn junction for $E_{\rm F}=1.1 \lambda_{\rm SO}$.}
\label{fano2d}
\end{figure}

In addition, we show the Fano factor in Fig. \ref{fano2d}.
%%%%%%%%
%%%%%%%%
The Fano factor (Fig. \ref{fano2d}) is basically given by the inverse of $G$ (Fig. \ref{pnp2d}): 
$F$ takes a small value for a resonant tunneling case.
In the long junction limit [Figs. \ref{fano2d}(c) and \ref{fano2d}(d)], $F$ of the npn junction tends to that of the pn junction [Fig. \ref{fano}(a)].
And also, in the insulating regime $(|E_{\rm F}-V|<\epsilon_1)$, $F$ converges to be unity. 
Namely, the Fano factor is interpreted to be unity for the insulating regime of pn junction.

\section{Summary}
\label{summary}

We have studied charge transport in the pn and npn junctions of silicene.
In silicene, the topological phase transition occurs by applying electric field owing to the buckling structure.
This transition affects the charge transport for the single-channel regime,
i.e., the resulting conductance is suppressed in the np regime for the TP/TP junction, 
while it is suppressed in the nn regime for the TP/NTP junction.
We have shown that this suppression originates from matching/mismatching of the spin and sublattice states of the incident and transmitted electrons.
Furthermore, the silicene pn junction has been shown to be a FET which conductance is almost quantized.
It is not the case in the graphene pn junction, which has no band gap.
The silicene junctions can be a potential new device controlled by two types of electric field $V$ and $E_z$.

\begin{acknowledgments}
This work is supported by the ``Topological Quantum
Phenomena" (No. 22103005) Grant-in Aid for Scientific
Research on Innovative Areas 
from the Ministry of Education,
Culture, Sports, Science and Technology (MEXT)
of Japan.
MH is supported by Grant-in Aid for Scientific Research No. 22740196.
\end{acknowledgments}

\if0
\appendix
\section{Calculation of conductance}
In this appendix, we explain about calculation of conductance of a silicene pn junction.
First we derive energy and wave function in the bulk silicene.
Hamiltonian of silicene is given by
\begin{align}
 H(\bm k) &= v_{\rm F} (k_x \tau_x - k_y \tau_y)
 + h_\tau(\bm k) \tau_z
 + \ell E_z \tau_z,
 \\
 h_\tau(\bm k) &= 
\bm g(\bm k) \cdot \bm \tau
=
-\lambda_{\rm SO} \tau_z + a \lambda_{\rm R}(k_x \tau_y - k_y \tau_x).
\end{align}
$g$-vector is given by 
\begin{align}
\bm g(\bm k) &= (g_x(\bm k),g_y(\bm k),g_z(\bm k)) 
\nonumber\\
&= (-a \lambda_{\rm R} k_y, a\lambda_{\rm R} k_x, -\lambda_{\rm SO}).
\end{align}
%
Eigenvalue of the spin part $h_\tau(\bm k)$ is obtained to be $\pm g(\bm k)$ with
\begin{align}
g(\bm k) &= \sqrt{ g_x^2(\bm k)+g_y^2(\bm k)+g_z^2(\bm k) }
\nonumber\\
 &=
\sqrt{\lambda_{\rm SO}^2 + a^2 \lambda_{\rm R}^2 k^2},
\end{align}
$k=(k_x^2+k_y^2)^{1/2}$, and the corresponding eigenvector is given by
\begin{align}
 \left| \bm k \pm \right\rangle_\tau
 = \frac{(g_x(\bm k) - i g_y(\bm k), \pm g(\bm k)-g_z(\bm k))^{\rm T}}{\sqrt{|k|^2+|k_x-ik_y|^2}},
\end{align}
\fi

\appendix

\section{Wave function in bulk silicene}
\label{wavefunction}

The low-energy Hamiltonian Eq. (\ref{hbulk}) is rewritten as
\begin{align}
 H(\bm k)
 &= v_{\rm F} (k_x \tau_x - k_y \tau_y \eta_z)
 + \left[ \bm g(\bm k) \cdot \bm \sigma \eta_z + \ell E_z  \right] \tau_z,
\end{align}
with $\bm g(\bm k) = (a \lambda_{\rm R} k_y, -a \lambda_{\rm R} k_x, -\lambda_{\rm SO})$.
We diagonalize this Hamiltonian sequentially, i.e., diagonalizing the spin part ($\sigma$) in the first step and sublattice pseudo spin ($\tau$) in the next step.
The eigenvalue of $\bm g(\bm k) \cdot \bm s$ is obtained to be $\pm g(\bm k)$ with
\begin{align}
 g(\bm k) = 
 \sqrt{g^2_x(\bm k) + g^2_y(\bm k) + g^2_z(\bm k)}
 =
\sqrt{a^2 \lambda_{\rm R}^2 k^2 + \lambda_{\rm SO}^2}.
\end{align}
The corresponding eigenvector $|\bm k \pm \rangle_\sigma$ is given by
\begin{align}
 |\bm k \pm \rangle_s 
\propto 
{[g_x(\bm k) -i g_y(\bm k) ]
\left| \uparrow \right\rangle + 
\left[\pm g(\bm k)-g_z(\bm k)
\right] 
\left| \downarrow \right\rangle
}.
\end{align}
As a result, the partially diagonalized Hamiltonian $H_{\pm}(\bm k)$ is given by
\begin{align}
 H_\pm(\bm k) = v_{\rm F} (k_x \tau_x - k_y \tau_y \eta_z)
 + [\pm g(\bm k) \eta_z + \ell E_z ] \tau_z.
\end{align}
Thus,
the energy spectrum is obtained to be $\pm E_{\pm}(\bm k)$ with
\begin{align}
 E_\pm^2(\bm k)
 &=
 v_{\rm F}^2 k^2 + \left[ \pm g(\bm k) + \ell E_z \right]^2.
\end{align}
%  with $k = (k_x^2+k_y^2)^{1/2}$.
% \begin{align}
% \eta(\bm k) = \mathrm{sgn} \left( E_z \right) \sqrt{a^2 \lambda_{\rm R}^2 k^2 + \lambda_{\rm SO}^2}.
% \end{align}
The corresponding eigenvector $\bm u_\pm(\bm k)$ is given by the direct product of the eigenvectors of the pseudo spin and the spin as
\begin{align}
 \bm u_\pm(\bm k)
 &= 
 \left|\bm k \pm \rangle_\tau
  |
\bm k \pm
  \right\rangle_\sigma,
\end{align}
with
\begin{align}
 \left| \bm k \pm \right \rangle_\tau
 & \propto 
 v_{\rm F}(k_x + i k_y \eta_z) \left| \mathrm A \right \rangle
 \nonumber\\ & \quad
 +
 [E_\pm(\bm k) - (\pm g(\bm k) \eta_z + \ell E_z)]
 \left| \mathrm B \right\rangle.
\end{align}

\section{Symmetry}
\label{symmetry}

In this Appendix, we show the symmetry of the conductance in the silicene junction.
Applying $\pi$-rotation along the $x$-axis, one obtains
\begin{align}
 \tau_x \sigma_x H(k_x,k_y)  \sigma_x \tau_x
 = H(k_x,-k_y) |_{E_z \to -E_z}.
\end{align}
The eigenvector $\bm u_\pm(\bm k)$ is transformed as
\begin{align}
 \tau_x \sigma_x \bm u_\pm(k_x,k_y)
 = \bm u_{\mp}(k_x,-k_y).
\end{align}
These lead to
\begin{align}
 r_{\alpha\beta} 
 = 
 r_{(-\alpha)(-\beta)}|_{k_y \to -k_y, E_z \to -E_z}.
 \label{sym1}
\end{align}
It follows that the charge conductance, which is obtained by the integral over $k_y$, $\alpha$, and $\beta$, is an even function of $E_z$.

Next, we show the relation of the conductance between the two valleys.
Applying unitary transformation $\tau_x \eta_x$, the Hamiltonian is transformed as
\begin{align}
 \tau_x \eta_x H(\bm k) \tau_x \eta_x
 = H(\bm k)|_{E_z \to - E_z}.
\end{align}
The eigenvector is also transformed as
\begin{align}
 \tau_x \bm u_\pm(\bm k) = \bm u_\pm(\bm k)|_{E_z \to -E_z, \eta_z \to -\eta_z}.
 \label{sym2}
\end{align}
% namely, $\rm K$ point with $E_z$ and $\rm K'$ point with $-E_z$ are the same.
From Eqs. (\ref{sym1}) and (\ref{sym2}), we conclude that the conductances contributed from K and K$'$ points are equivalent to each other.

\bibliography{pn}

\clearpage

\end{document}